\documentstyle[sprocl,epsf]{article}

\def\Pom{{\bf I\!P}}
\def\lsim{\mathrel{\rlap{\lower4pt\hbox{\hskip1pt$\sim$}}
    \raise1pt\hbox{$<$}}}         %less than or approx. symbol
\def\gsim{\mathrel{\rlap{\lower4pt\hbox{\hskip1pt$\sim$}}
    \raise1pt\hbox{$>$}}}         %greater than or approx. symbol

\def\beq{\begin{equation}}
\def\eeq{\end{equation}}
\def\bea{\begin{eqnarray}}
\def\eea{\end{eqnarray}}
\begin{document}
\rightline{{\bf\em FZ-IKP(TH)-1998-6}}
\vspace{1.0cm}
\title{FORWARD CONE IN EXCLUSIVE HEAVY VECTOR MESON PRODUCTION
 \footnote {Talk at 6th International Workshop on
 Deep Inelastic Scattering and QCD (DIS'98), Brussels, April 1998}}

%\author{A. B. AUTHOR, C. D. AUTHOR}

%\address{World Scientific Publishing Co, 1060 Main Street, 
%River Edge,\\ NJ 07661, USA\\E-mail: wspc@wspc.com} 

\author{V. R. ZOLLER}

\address{Institute for  Theoretical and Experimental Physics,\\
Moscow 117218, Russia\\E-mail: zoller@heron.itep.ru}
%%%%%%%%%%%%%%%%%%%%%%%%%%%%%%%%%%%%%%%%%%%%%%%%%%%%%%%%%%%%%%
% You may repeat \author \address as often as necessary      %
%%%%%%%%%%%%%%%%%%%%%%%%%%%%%%%%%%%%%%%%%%%%%%%%%%%%%%%%%%%%%%

\maketitle\abstracts{The 
color dipole gBFKL phenomenology of a
diffraction cone for photo- and electroproduction
$\gamma^{*}N \rightarrow VN$ of heavy vector mesons
(charmonium \& bottonium) at HERA and in fixed target
experiments is presented. A substantial shrinkage of the
diffraction cone
from the CERN/FNAL to the HERA range of c.m.s. energy $W$ is predicted.
 The
$Q^{2}$-controlled selectivity to the color dipole size
(scanning phenomenon) is shown to lead to a decrease of the
diffraction slope with $Q^{2}$ (which is supported by the
available experimental data).  An approximate
flavor independence of the diffraction slope in the scaling
variable $Q^{2}+m_{V}^{2}$ is shown to hold.    }
%........................................................
\section{Introduction}
%........................................................
 I report on the color
dipole description of the forward diffraction slope 
$B(\gamma^{*}\rightarrow V)$ in exclusive diffractive DIS \cite{ONE}. 
We use our early results for the energy dependence of the forward 
cone in color dipole scattering \cite{NZZslope}, obtained from the
solution of the running gBFKL equation for the diffraction slope
\cite{NZZspectrum}.  The crucial point is that breaking of the scale
invariance by asymptotic freedom (running  $\alpha_{S}(r)$) and finite
propagation radius $R_{c}$ for perturbative gluons,
alters  the BFKL pomeron from a 
{\sl fixed cut} in the scaling approximation ($\alpha_{S}={\rm const}$)
 to a series of moving 
poles for the running gBFKL pomeron \cite{NZZspectrum,NZZRegge} (for early
quasiclassical analysis see also \cite{Lipatov}).  
The two major points about $B(\gamma^{*}\rightarrow V)$ which we focus here 
are: i) a prediction of a substantial Regge shrinkage of the diffraction slope 
for the running gBFKL; ii) a decrease of the slope 
 with $Q^2$ and/or mass of the vector meson
because of a shrinkage of the photon quantified by the scanning radius
\cite{KNNZ93} 
%
%
% -----------------------------------------------
$r_{S} \approx {A/ \sqrt{m_{V}^{2}+Q^{2}}}$.
% -----------------------------------------------
%
%
\section{Color dipole factorization for the diffraction cone}
To the leading $\log(1/x_{eff})$ approximation
the amplitude for real (virtual) photoproduction
of vector mesons with the momentum transfer $\vec{q}$ can be
cast in the color dipole factorized form
%
%
% -----------------------------------------------
\beq
{\cal M}(\gamma^{*}\rightarrow V;\xi,Q^{2},\vec{q})=
\int\limits_{0}^{1} dz\int d^{2}\vec{r}
\Psi_{V}^{*}(r,z)
{\cal A}(x_{eff},r,z,\vec{q})
\Psi_{\gamma^{*}}(r,z)\,,
\label{eq:2.2}
\eeq
% ------------------------------------------------
%
%
%
%
%-------------------------------------------------
where  $\Psi_{\gamma^{*}}(\vec{r},z)$ and
$\Psi_{V}(\vec{r},z)$ are the probability amplitudes
to find the color dipole  $\vec r$
in the photon and vector meson, respectively,
and ${\cal A}(x_{eff},r,z,\vec{q})$ is the imaginary part of the 
 amplitude for the color dipole
elastic scattering  on the target nucleon,
${\cal A}(x_{eff},r,z,0)=\sigma(x_{eff},r)$,where
$x_{eff}=(Q^2+m^2_V)/W^2$, $\xi=\log(1/x_{eff})$.

We define the diffraction slope by 
$B=\left.- 
{d\log(d\sigma/dq^2)/ dq^{2}}\right|_{q=0}$.
%------------------------------------------------
%
The representation for ${\cal A}(x_{eff},r,\vec{q})$ in terms
of the gluon density matrix is
%
%
%------------------------------------------------
\bea
{\cal A}(x_{eff},r,\vec{q})=
{4\pi \over 3}\int {d^{2}\vec{k}\over k^{4}}
\, \alpha_{S}(\kappa^{2})
[J_{0}({1\over 2}qr)-J_{0}(kr)]
{\cal F}(x,\vec{k},\vec{q})\,,
\label{eq:2.7}
\eea
%------------------------------------------------
%
%
where $J_{0}(y)$ is the Bessel function.
The gluon density matrix
${\cal F}(x,\vec{k},\vec{q})$
is proportional to the imaginary part of the non-forward gluon-nucleon
scattering amplitude
%
%
%------------------------------------------------
\beq
{\cal F}(x_{eff},\vec{q},\vec{k})=
{4 \over \pi}
\int d^{2}\vec{k}_{1}
{\cal T}(x_{eff},\vec{k},\vec{q},\vec{k}_{1})
\alpha_{S}(k_{1}^{2})V_p(\vec{k}_1,\vec{q})\,,
\label{eq:4.1}
\eeq
%------------------------------------------------
%
%
where the proton vertex function $V_p$
 is represented in terms of 
the single- and two-quark form factors of the proton 
$V_{p}(\vec{k}_{1},\vec{q})=
G_{1}(q^{2})-G_{2}(\vec{k}_{1},\vec{q})$ and
${\cal T}(x,\vec{k},\vec{q},\vec{k}_{1})$
stands for the propagation function of two $t$-channel gluons.
Splitting the color dipole vertex function 
$V_{d}(q,r)=[J_{0}({1\over 2}qr)-J_{0}(kr)]$ into two pieces,
 we obtain a useful
decomposition
%
%
%------------------------------------------------
%
%
%------------------------------------------------
\bea
{\cal A}(x_{eff},r,\vec{q})=
{4\pi \over 3}[J_{0}({1\over 2}qr)-1]\int {d^{2}\vec{k}\over k^{4}}
\, \alpha_{S}(\kappa^{2})
{\cal F}(x,\vec{k},\vec{q})\nonumber\\
+
{16 \over 9}\int {d^{2}\vec{k}\over k^4}\, \alpha_{S}(\kappa^{2})
[1-J_{0}(kr)]
\int d^{2}\vec{k}_{1}
{\cal T}(x,\vec{k},\vec{q},
\vec{k}_{1})
\alpha_{S}(k_{1}^{2})
V_p(\vec{k}_{1},\vec{q}) \, ,
\label{eq:4.4}
\eea
%------------------------------------------------
%
%
which  illustrates how the
three relevant size parameters in the problem give
rise to the three major components of the diffraction slope.
The $q$ dependence coming from 
the proton vertex function
$V_{p}(\vec{k}_{1},\vec{q})$
 is controlled by the proton
size $R_N$.  The  color dipole vertex
function $V_{d}=J_{0}({1\over 2}qr)-1$ gives a
 geometric contribution $\Delta B_d(r)=r^2/8$.
The BFKL diffusion of gluons
in the impact parameter plane gives rise to the asymptotic Regge growth 
of the diffraction slope \cite{NZZslope} such that for all $\vec k$, $\vec k_1$
one has
${\partial {\cal T}/ \partial q^{2}}|_{q^{2}=0}
=
-[\alpha_{\Pom}'\xi+{\cal O}(R_{c}^{2})]
{\cal T}$.
Besides, the BFKL diffusion results in  azimuthal
decorrelation of  $\vec k$, $\vec k_1$. This property of the BFKL kernel
ensures the additivity of the beam, target and exchange contributions to 
$B=\Delta B_{d}(r)+\Delta B_N+\Delta B_{Regge}$.

 For large dipoles, $r \gsim R_{c}$,
  one recovers a sort of
additive quark model
%------------------------------------------------
\beq
B(\xi,r)=
\frac{1}{8}r^{2}+\frac{1}{3}R_{N}^{2}+
2\alpha_{\Pom}'\xi + {\cal O}(R_{c}^{2})\, ,
\label{eq:4.13}
\eeq
%------------------------------------------------
%
%
For small dipoles, $r\lsim R_{c}$,
 the   $\gamma V$ vertex contribution is of the form 
($A_{\sigma}\simeq 10$)
\beq
\Delta B_{d}(r) = {R_{c}^{2} \over 8}{G(x_{eff},{A_{\sigma}/ R_{c}^{2}})
\over G(x_{eff},{A_{\sigma}/r^{2}})}
\label{eq:3.28}
\eeq
which vanishes at small $r$ only because of scaling
 violation in the gluon density
$G(x_{eff},{A_{\sigma}/r^{2}})$,
 much slower than the geometrical term $r^2/8$.
 At moderately small $x$ values,
 the ratio of gluon densities  has a substantial $x_{eff}$ dependence
and $\Delta B_{d}$ affects substantially the energy dependence of $B$.
\begin{figure}
\epsfxsize=0.8\hsize
\epsfbox{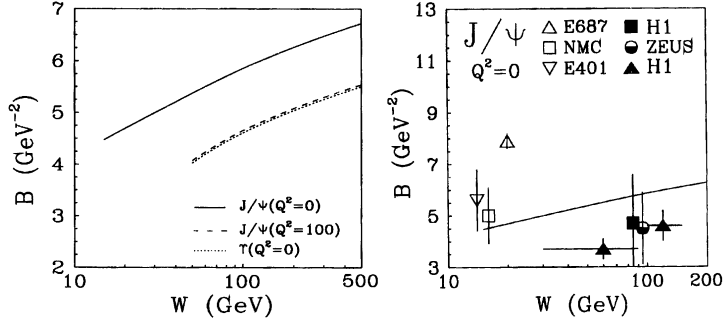}
\caption{{\rm left box}~- The color dipole model predictions for the
c.m.s. energy dependence of the diffraction slope
for real photoproduction of the $J/\Psi$ and $\Upsilon$
and for the $J/\Psi$ electroproduction at $Q^{2}=100$\,GeV$^{2}$;
{\rm right box}~- Comparison of the color dipole model prediction
for c.m.s. energy $W$ dependence of the
diffraction slope for photoproduction of the
$J/\Psi$ with the data.}
\label{fig1}
\end{figure}

\section{Predictions vs. data}
Because of the scanning property, 
$${\cal M}(\gamma^*\to V;\xi,Q^2,\vec q)\propto
{\cal A}(x_{eff},r_S,z\simeq 1/2,\vec q)\,,$$
 The contribution to the diffraction slope
from the term $\Delta B_{d}(r_{S})$  decreases
with the decreasing scanning radius $r_{S}$, i.e.,
with rising $Q^{2}$ \cite{NZZslope}. At fixed energy
$W$ the value of $x_{eff}$ rises and the rapidity 
$\xi$
decreases which also diminishes the diffraction slope
because the Regge term becomes smaller.

 The predicted energy dependence of
the diffraction slope for real photoproduction of the
$J/\Psi$ and $\Upsilon$ is presented in Fig.~1.
 The diffraction slope
 grows by $\sim 1.5$\,GeV$^{-2}$
when $W$ grows by one order in magnitude from the fixed
target up to the HERA energy.
This corresponds to the effective shrinkage rate
$\alpha_{eff}' \approx 0.15$\,GeV$^{-2}$.
 The diffraction slope
for the $J/\Psi$ production at $Q^{2}=100$\,GeV$^{2}$
nearly coincides with that for real photoproduction
of $\Upsilon$
 because the scanning radii  
$r_{S}$ for the two reactions are very
close to each other.
The experimental
situation in the
heavy quarkonium sector is summarized in Fig.~3.
 At the both fixed
target \cite{E687Psi',E401Psi,NMCPsi} and HERA energy
\cite{ZEUSPsi,ZEUSjp97,H1Psi,H1PsiW96}
 the error bars are too big for the
definitive conclusions on the presence and/or lack of the
shrinkage of the diffraction cone to be drawn.
The admixture of inelastic events makes 
it difficult to compare different experiments.
The slope for $J/\Psi$ as  measured at HERA is too small,
 even smaller than 
the proton electromagnetic size contribution to $B$, $\Delta B_N
=\frac{1}{3}R_{N}^{2}\simeq 5.8\,GeV^{-2}$. The resolution of the paradox
requires a better understanding of the role
 of secondary reggeons. 
\begin{figure}
\epsfxsize=0.7\hsize
\epsfbox{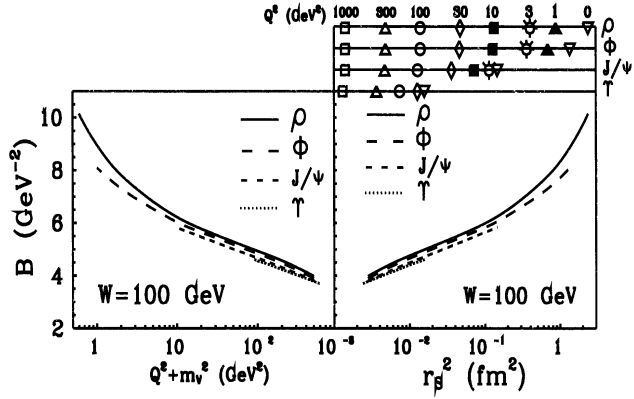}
\caption{~- The color dipole model
predictions for the
diffraction slope in production of different
vector mesons as a function of the scaling variable
$m_{V}^{2}+Q^{2}$ (left box) and the scanning radius $r_{S}$
(right box) at fixed
c.m.s. energy $W=100$\,GeV. 
The scales of $Q^{2}$ on the top of the right box 
show the values of $Q^{2}$ which
correspond to the scanning radii shown on the bottom axis.}
\label{fig2}
\end{figure}
The flavor symmetry properties of the diffraction cone can be 
seen also in  Fig.~2.
 The curves for
$B(\gamma^{*}\rightarrow V)$ of all the vector
mesons do converge together  as a
function of $Q^{2}+m_{V}^{2}$.
In Fig.~2b the same results are presented as a
function of the scanning radius $r_{S}$ with $A=6$.
We  suggest useful empirical parameterizations for the
diffraction slope.
For production of heavy quarkonia the
$Q^{2}$ dependence of the diffraction slope at $W=100$\,GeV
and in the considered range of $Q^{2}\lsim 500$\,GeV$^{2}$ can
be approximated by
%
%
%------------------------------------------------
\beq
B(\gamma^{*}\rightarrow V) \approx \beta_{0} - \beta_{1}
\log\left({Q^{2}+m_{V}^{2} \over m_{J/\Psi}^{2}}\right)
\label{eq:8.1}
\eeq
%------------------------------------------------
%
%
with the slope $\beta_{1} \approx 1.1$\,GeV$^{-2}$ and the
constant $\beta_{0}\approx 5.8$\,GeV$^{-2}$.
The constant $\beta_{0}$ is subject to the choice of the $t$
range ($|t|\simeq 0.1\,{\rm GeV^2}$),
 it is the value of the parameter $\beta_{1}$ which is 
closely related to the gBFKL dynamics.
For 
virtual production of $J/\Psi$ 
the H1
\cite{H1PsiQ2} reported:
$B=3.8\pm 1.2(st)^{+2.0}_{-1.6}(syst)$ ${\rm GeV}^{-2}$
 at $ <Q^{2}>=18\,{\rm GeV}^{2}$, $W=90\,{\rm GeV} $.
 The ZEUS collaboration \cite{ZEUSPsi97Q2} presented  
the value
$B=4.5\pm 0.8(st)\pm 1.0(syst)$\,GeV$^{-2}$ 
at $Q^{2}=$\,6 GeV$^{2}$.
Predicted
decrease of the diffraction slope from $Q^{2}=0$ to
$Q^{2}=18$\,GeV$^{2}$ by mere $\approx 0.5$\,GeV$^{-2}$,
is too small an effect to be seen at the present experimental
accuracy.

%\begin{figure}[t]
%\rule{5cm}{0.2mm}\hfill\rule{5cm}{0.2mm}
%\vskip 2.5cm
%\rule{5cm}{0.2mm}\hfill\rule{5cm}{0.2mm}
%%\psfig{figure=eig0126.ps,height=1.5in}
%\caption{A generalized cactus tree: the confluent
%transfer-matrix $S$ transforms the state function $f(x)$ and
%$f(z)$ into $f(x)$.  \label{fig:radish}}
%\end{figure}

\end{document}